\documentclass[prl,twocolumn,floatfix,showpacs,10pt,a4paper]{revtex4}
\usepackage{amssymb}
\usepackage{amsmath}
\usepackage[dvipdfm]{graphicx}
\usepackage{subfigure}
\usepackage{times}\usepackage[latin1]{inputenc}
\usepackage[all]{xy}
\usepackage{graphicx}
\newcommand{\bb}{\bibitem}
\newcommand{\be}{\begin{equation}}
\newcommand{\ee}{\end{equation}}
\newcommand{\ben}{\begin{eqnarray}}
\newcommand{\een}{\end{eqnarray}}
\newcommand{\bes}{\begin{subequations}}
\newcommand{\ees}{\end{subequations}}

\begin{document}
\title{Solitons with Cubic and Quintic Nonlinearities Modulated in Space and Time}
\author{A.T. Avelar$\,^a$, D. Bazeia$^b$, and W.B. Cardoso$^a$}
\affiliation{$^a$Instituto de F\'\i sica, Universidade Federal de Goi\'as, 74.001-970, Goi\^ania
Goi\'as, Brazil\\$^b$Departamento de F\'\i sica, Universidade Federal da Para\'\i ba, 58.051-970, Jo\~ao
Pessoa Para\'\i ba, Brazil}
\begin{abstract}
This work deals with soliton solutions of the nonlinear Schroedinger equation with cubic and quintic nonlinearities. We extend the procedure put forward in a recent Letter [J. Belmonte-Beitia et al., Phys. Rev. Lett. 100, 164102 (2008)], and we solve the equation in the presence of linear background, and cubic and quintic interactions which are modulated in space and time. As a result, we show how a simple parameter can be used to generate brightlike or darklike localized nonlinear waves which oscillate in several distinct ways, driven by the space and time dependence of the parameters that control the trapping potential, and the cubic and quintic nonlinearities.
\end{abstract}

\maketitle

The nonlinear Schroedinger equation (NLSE) has been studied in a diversity of situations. The nonlinear interactions are usually of cubic nature, but there are systems which engender cubic and quintic (CQ) nonlinearities. The case of CQ nonlinearities opens new possibilities, and there are many interesting applications, specially in nonlinear optics, in fibers where the CQ nonlinearities may be used for instance to describe pulse propagation in double-doped optical fibers, when the type of dopant varies along the fiber, with the value and sign of the cubic and quintic parameters that control the nonlinearities being adjusted by properly choosing the characteristics of the two dopants \cite{GH}. There are other applications, and here we quote the Bose-Einstein condensates (BEC), where a diversity of nonlinearities may appear driven by controlled optical interactions \cite{Theis}.

A particularly important scenario for solitons concerns the NLSE in the case of a single spatial dimension with a linear term and cubic nonlinearity. 
However, we can change the linear parameter to make it to trap the system. In this case it is named the Gross-Pitaevskii equation \cite{DP}, and the linear parameter $v$ is now modified to $v(x)$, with explicit dependence on the spatial coordinate $x$. Usually, the potential $v(x)$ is a background potential, and has the form of the harmonic potential, with the purpose of trapping the system in a finite region in space, or some spatially periodically oscillating pattern, if the purpose is to entrap the system into a periodic lattice. This equation has gained further importance recently, mainly because of its direct application to the study of BECs \cite{B1}, in fibers and in photonic crystals and other periodic systems \cite{B2,B3}. Other applications include the study of nonlinear tunneling of spatial and temporal optical solitons in optical organic materials of practical use to ultrafast photonic technologies \cite{SC}.

The pioneer work by Serkin and Hasegawa (SH) \cite{SH} has introduced an interesting procedure to deal with nonautonomous NLSE. It is based on a similarity transformation, which transforms nonautonomous NLSE equation into stationary equation which is easier to solve. The SH procedure has inspired several investigations \cite{other}. In particular, in the recent work by Belmonte-Beitia, P\'erez-Garc\'\i a, Vekslerchik, and Konotop (BPVK) \cite{BB} one deals with the equation 
\begin{equation}\label{NLSB}
i\frac{\partial\Psi}{\partial t}=- \frac{\partial^{2}\Psi}{\partial x^{2}}
+v(x,t)\Psi +g(x,t)|\Psi|^{2}\Psi.
\end{equation}
Here $v(x,t)$ and $g(x,t)$ may now vary, being functions with both space and time dependence, the first one being the trapping potential, and the second one describing the cubic nonlinearity. In the present work we shall follow the BPVK procedure, which is well explained in \cite{BB} and so we do not review it here. Instead, we extend the method to another problem, focusing our attention on the CQ nonlinear equation 
\begin{equation}
i\frac{\partial\Psi}{\partial t}\!=\!-\frac{\partial^{2}\Psi}{\partial x^{2}}+v\!(x,t)\Psi\!+\!g_3\!(x,t)|\Psi|^{2}\Psi\!+\!g_5\!(x,t)|\Psi|^{4}\Psi,
\label{CQNLS}
\end{equation}
where $v(x,t)$ is the trapping potential, and $g_3(x,t)$ and $g_5(x,t)$ control the nonlinear cubic and quintic interactions, respectively. The validity
of the one dimensional approach is given explicitly in \cite{BB}, and the quintic term which we have added in \eqref{CQNLS} may simulate three-body collisions and/or deviation of the trapped condensate from the one-dimensionality \cite{KVM}. The trapping potential and nonlinearities to be used below are typical of BECs, and the results obtained may stimulate new experiments in the field.

In this work, we provide a way of making the BPVK procedure work in this scenario, involving the presence of CQ nonlinearities. The result allows one to obtain explicit solutions for some specific choices of parameters, leading us to interesting localized solutions of the bright or dark type, depending on the vanishing or not of the eigenvalue of the associated stationary nonlinear equation, as we explain below. 

The idea is to write the solution of \eqref{CQNLS} as
\begin{equation}
\Psi (x,t)=\rho(x,t)e^{i\phi(x,t)}\Phi(\zeta(x,t)),\label{TS}
\end{equation}
in order to rewrite \eqref{CQNLS} into the stationary equation 
\begin{equation}
\mu\Phi(\zeta )=-\frac{d^{2}\Phi(\zeta)}{d\zeta^{2}}
+G_3|\Phi(\zeta)|^{2}\Phi(\zeta)+G_5|\Phi(\zeta )|^{4}\Phi(\zeta),\label{Q}
\end{equation}
where $\mu$ is the eigenvalue of the nonlinear equation above, and $G_3$ and $G_5$ are real constants which control the CQ nonlinearities. The substitution of \eqref{TS} into \eqref{CQNLS} leads to (\ref{Q}), but now we have to have 
\bes\label{nl1}\ben
&&\rho\frac{\partial\rho}{\partial t}+\frac{\partial(\rho^{2}(\partial\phi/\partial x))}{\partial x}=0,
\\
&&\frac{\partial\zeta}{\partial t}+2\frac{\partial\phi}{\partial x}\frac{\partial\zeta}{\partial x}=0,\;\;\;\;\;
\frac{\partial\left(\rho^{2}(\partial\zeta/\partial x)\right)}{\partial x}=0.
\een\ees

We can introduce a new function $\xi(x,t)$ such that $\zeta(x,t)=F(\xi(x,t))$. In this case, we write $\xi (x,t)=\gamma(t)x+\delta(t)$. This choice is interesting because it allows to determine the width of the localized solution in the form $1/\gamma (t)$, and its center of mass position as
$-\delta (t)/\gamma (t)$. With the Eqs.~\eqref{nl1} we can obtain the new set of equations 
\bes\label{nl2}\ben
\rho (x,t)&=&\left({\gamma}/{(\partial F/\partial\xi)}\right)^{1/2},
\\
\phi(x,t)&=&-\frac{(\partial\gamma/\partial t)}{4\gamma}x^{2}-\frac{(\partial\delta/\partial t)}{2\gamma}x+a(t),\label{q}
\\
v(x,t)&=&\frac{1}{\rho}\frac{\partial^{2}\rho}{\partial x^{2}}-\frac{\partial\phi}{\partial t}-\left(\frac{\partial\phi}{\partial x}\right)^{2}-
\frac{\mu\gamma^{4}}{\rho^{4}},\label{vx}
\\
g_3(x,t)&=&G_3{\gamma^{4}}{\rho ^{-6}},
\\
g_5(x,t)&=&G_5G_3^{-1}\rho^{-2}g_3(x,t)=G_5{\gamma^{4}}{\rho^{-8}},\label{nl2b}
\een\ees
where $a(t)$ is an arbitrary function of time. The choice of $F(\xi)$ is to be done in a way such that we get finite energy solutions of the CQ nonlinear equation. 

This is the general procedure, and we note from \eqref{nl2b} that if we take the limit $G_5\to0$ we get to $g_5(x,t)\to0$ and this leads us back to the case set forward in \cite{BB}. This result is robust. In fact, we have shown that the procedure can be extended to other nonlinearities: if in \eqref{CQNLS} we add the new term $g_7(x,t)|\Psi|^6\Psi$, etc, then in \eqref{Q} we should add $G_7|\Phi|^6\Phi$, etc, and after \eqref{nl2b} we should include the new expression
$g_7(x,t)\!\!=\!\!G_7G_5^{-1}\!\rho^{-2}g_5(x,t)\!\!=\!\!G_7G_3^{-1}\!\rho^{-4}g_3(x,t)\!\!=\!\!G_7{\gamma^4}{\rho^{-10}}$, and so on. We will deal with the more general case in another work \cite{abc1}, and below we focus attention on the important case of CQ nonlinearities.

Up the here, we have been very general. However, to illustrate the procedure with examples of interest, let us focus our attention into the case of specific nonlinearities. This is a nontrivial task, but we know that in the case of BECs with controlled optical interactions, we can have a diversity of nonlinearities \cite{GH,Theis,DP,B1,B2,B3,SC,SH,other,BB,KVM}. Thus, in the present work we suppose that the cubic nonlinearity is given explicitly by  
\be\label{g3}
g_3(x,t)=\gamma e^{\xi^2/b^2},
\ee
where $b$ is real parameter which controls the behavior of the cubic nonlinearity -- we will use $b=8$ where required. With this choice we have
\be
\rho(x,t)=G_3^{1/6}\gamma^{1/2}e^{-\xi^2/6b^2}.\label{ro}
\ee
The quintic nonlinearity is now controlled by
\be
g_5(x,t)=G_5G_3^{-4/3}e^{4\xi^2/3b^2},
\ee
and the potential given by Eq.~\eqref{vx} has to have the form
\ben
v(x,t)\!=\!\omega^{2}x^{2}\!\!+\!\!f_1x\!+\!\!f_2\!-\!\mu \gamma^2 G_3^{-2/3} e^{2\xi^2/3b^2},
\een
and $\omega,$ $f_1$, and $f_2$, are time-dependent functions such that
\bes\label{13}\ben
\omega^{2}(t)=\gamma^{4}+\frac{1}{4\gamma}\frac{d^{2}\gamma}{dt^{2}}-\frac{1}{2\gamma^{2}}\left(\frac{d\gamma}{dt}\right)^{2},\label{13a}
\\
f_1(t)=2\gamma^{3}\delta+\frac{1}{2\gamma}\frac{d^{2}\delta}{dt^{2}}-\frac{1}{\gamma^{2}}\frac{d\gamma}{dt}\frac{d\delta}{dt},\label{13b}
\\
f_2(t)=\gamma^{2}+\gamma^{2}\delta^{2}-\frac{1}{4\gamma^{2}}\left(\frac{d\delta}{dt}\right)^{2}-\frac{da}{dt}.\label{13c}
\een\ees
   
We define $\chi=1/\gamma$ in order to rewrite \eqref{13a} in the form of the Ermakov-Pinney equation \cite{BB,PG}
\be
\frac{d^{2}\chi }{dt^{2}}+4\omega ^{2}(t)\chi =\frac{4}{\chi ^{3}}.\label{Ermakov}
\ee
This equation has solution for $\gamma$ in the form
\be
\gamma(t)=\left[ 2y_{1}^{2}(t)+2y_{2}^{2}(t)/W^{2}\right]^{-1/2}.
\ee
Here $W$ is the Wronskian of the two linearly independent solutions $y_{1}$ e $y_{2}$ of the Mathieu equation
$({d^{2}y}/{dt^{2}})\!+\!4\omega ^{2}(t)y\!=\!0$. See Ref.~{\cite{PG}} for more details on this. 

We now choose
\be
\omega^{2}(t)=1+\varepsilon\cos(\omega _{0}t),\label{omega}
\ee
in order to obtain analytical solutions. The above results are general and can be used to investigate explicit examples, as we consider below.  
Before searching for explicit solutions, however, let us remark that investigations concerning specific criteria for adiabaticity of nonlinear wave equations and of the soliton solutions to such equations, and issues related to parametric amplification of elementary excitations due to the periodic modulation of the trapping potential and nonlinearities were studied before in \cite{KM} and \cite{SLV}, respectively.

An important issue which appears in the case of CQ nonlinearities is that we can solve Eq.~\eqref{Q} for bright and dark solitons, depending of the values of the parameters that control the nonlinearities in a specific system. As the first example, let us consider
the simpler case in which the stationary Eq.~\eqref{Q} has vanishing eigenvalue, $\mu=0$. In this case, if we choose the cubic and quintic parameters as $G_3=2$ and $G_5=-3$, we can write Eq.~\eqref{Q} in the form 
\be
\frac{d^{2}\Phi(\zeta)}{d\zeta^{2}}=2\Phi^3(\zeta)-3\Phi^5(\zeta).\label{BS}
\ee
The solution is given by $\Phi(\zeta )={1}/{\sqrt{1+\zeta^{2}}}$. It has the bell shape form, and it is of the bright type, as we plot in Fig.~1(a). To have the correct boundary condition, we have to set $\zeta\to\pm\infty$ as $x\to\pm\infty$, and so we have to choose $\zeta(x,t)=F(\xi(x,t))$ properly in order to make the solution to behave according to the required boundary conditions. According to Eqs.~\eqref{nl2}, we choose $F(\xi)=G_3^{-1/3}\int d\xi e^{\xi^2/3b^2}$. We note that the bright solution is localized, but it is thicker then the hyperbolic secant, the standard bright soliton, which is also plotted in Fig.~1(a), for comparison.

As we know, the form found above for the bright soliton can also be obtained in models of relativistic scalar fields, but this will be reported in another work \cite{abc2}. We will get into this following the lines of \cite{ablm}, which suggests several distinct ways of modifying the bell shaped form of brightlike solutions, an issue of direct interest for practical applications, as we will illustrate in \cite{abc1}.

To get to the darklike solution, let us now choose a nonvanishing eigenvalue, $\mu\neq0.$ In order to get explicitly simple analytical solution we consider, for instance, the case $\mu=3$, $G_3=6$, and $G_5=-3$. In this case, Eq.~\eqref{Q} changes to
\be
\frac{d^{2}\Phi(\zeta)}{d\zeta^{2}}=-3\Phi(\zeta)+6\Phi^3(\zeta)-3\Phi^5(\zeta).\label{DS}
\ee
The solution is given by $\Phi(\zeta)={\zeta}/{\sqrt{1+\zeta^{2}}}$. It has the form of a kink, as we show in Fig.~1(b), and so $|\Phi(\zeta)|^{2}$ is now of the dark type. It is interesting to note that this form of solution has also appeared in high energy physics, as reported before in \cite{blm}. It is thicker then the standard darklike soliton which is described by the hyperbolic tangent, which we also plot in Fig.~1(b), for comparison.

\begin{figure}[t]
\includegraphics[width=3.4cm]{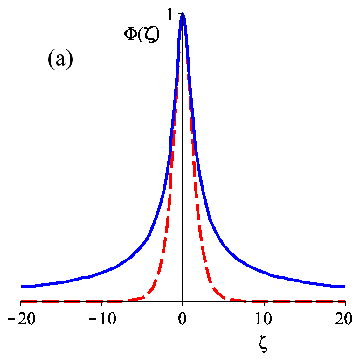}\hspace{1.0cm}
\includegraphics[width=3.4cm]{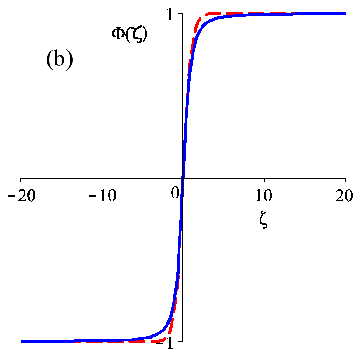}
\caption{(Color on line) Plots of $\Phi(\zeta)$ for the bright soliton (a) (blue, solid line) and the ${\rm sech}(\zeta)$ (red, dashed line), and for the dark soliton (b) (blue, solid line) and the $\tanh(\zeta)$ (red, dashed line).}
\end{figure}

\begin{figure}[b]
\includegraphics[width=3.5cm]{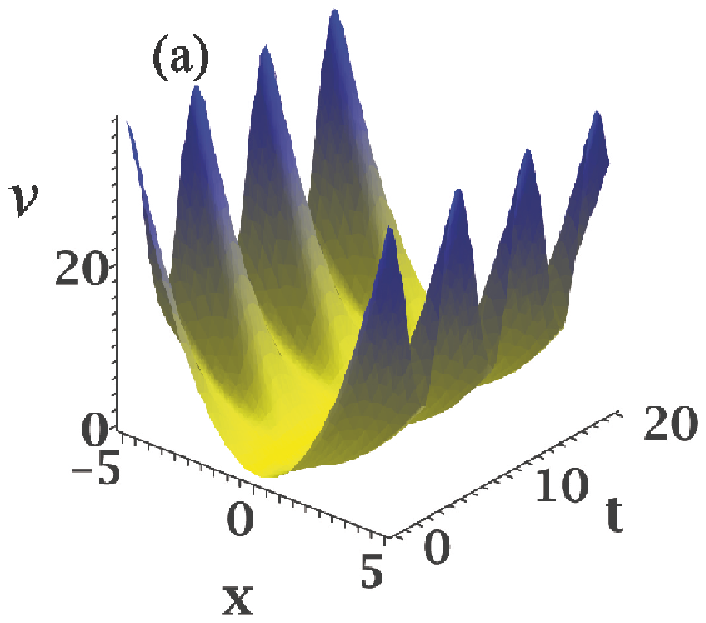}\hspace{0.5cm}
\includegraphics[width=3.5cm]{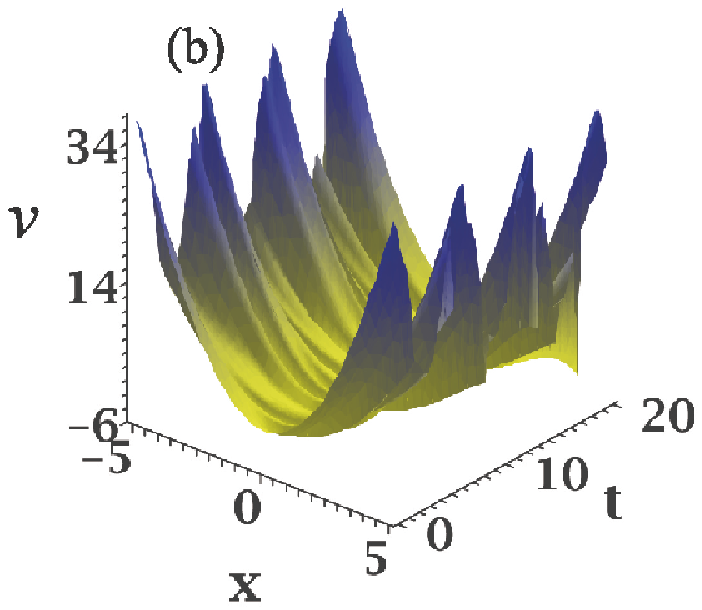}
\caption{(Color on line) Plots of the trapping potential $v(x,t)$ given by \eqref{v1} for the resonant brightlike case (a) and by \eqref{v2} for the resonant darklike case (b), in the range $-3<x<3$ for the time evolution in the interval $0<t<20$.}
\label{nlpot}
\end{figure}

Equation \eqref{BS} leads to the brightlike solutions. In this case the wave function which solves Eq.~\eqref{CQNLS} acquires the form
\be
\Psi(x,t)=\frac{2^{1/6}\gamma^{1/2}e^{i\phi}e^{-\xi^2/6b^2}}{\sqrt{1+\zeta^{2}}},\label{psibr}
\ee
where $\phi=\phi(x,t)$ is real, obtained via the Eq.~\eqref{q}.

A specific solution can be constructed, which leads to resonant solitons.
It is given by $\delta (t)=0$, $\mu =0$ and $a(t)=\int\gamma^{2}(t)dt$. In this case we get
\bes\label{nlp}\ben
v(x,t)\!&=&\!\omega ^{2}(t)x^{2},\label{v1}
\\
g_5(x,t)\!&=&\!-3(2)^{-4/3}e^{4\xi^2/3b^2},
\een\ees
with $g_3(x,t)$ given by \eqref{g3}. In Fig.~2(a) we plot the trapping potential \eqref{v1} in the standard form, to show how it behaves as a function of space and time. This is to be compared with the trapping potential of the other case, for $\mu\neq0$, where the dark solitons appear.

Solutions of the brightlike form nicely appear when one adequately chooses both $\varepsilon$ and $\omega_{0}$ in Eq.~\eqref{omega}. In Fig.~3, we show the resonant and breathing solitons for the specific choices of parameters $\epsilon=0.5$ and $\omega_{0}=1$, and $\epsilon=0,$ respectively. We can also get quasiperiodic and moving solitons, but we will leave this to the longer work \cite{abc1}.

\begin{figure}[t]
\includegraphics[width=4.1cm]{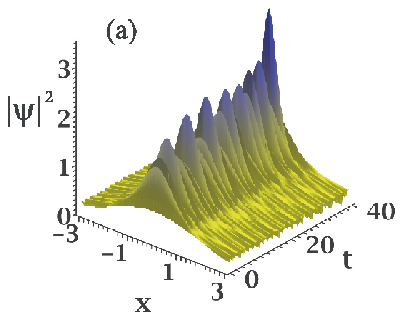}\hspace{0.2cm}
\includegraphics[width=4.1cm]{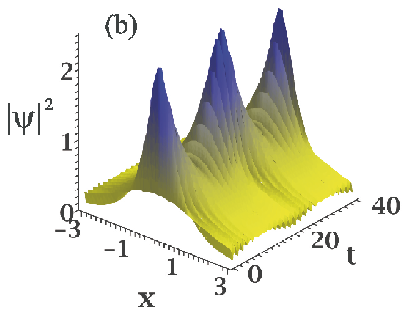}
\caption{(Color on line) Plots of $|\Psi(x,t)|^2$ for the resonant bright soliton for $\epsilon=0.5$ and $\omega_{0}=1$ (a), and of the breathing bright soliton, for $\epsilon=0$ (b). Initial data for Eq.~\eqref{Ermakov} are $\chi(0)=\sqrt{2}$ and ${d\chi}/{dt}(0)=0$.}
\label{bsrs}
\end{figure}

As we have already seen, the Eq.~\eqref{DS} gives the darklike solutions. In this case, the Eq.~\eqref{TS} takes the form
\be
\Psi(x,t)=\frac{6^{1/6}\gamma^{1/2}e^{i\phi} \zeta e^{-\xi^2/6b^2}}{\sqrt{1+\zeta^{2}}}.\label{psids}
\ee
If we take $\delta(t)=0$, $\mu\neq0$ e $a(t)=\int\gamma^2(t)dt$, we get
\bes\ben
v(x,t)\!\!&=&\!\!\omega^{2}(t)x^{2}\!\!-\!\mu\, 6^{-2/3}\gamma^2 e^{2\xi^2/3b^2} ,\label{v2}
\\
g_5(x,t)\!\!&=&\!\!-\frac12\,6^{-1/3} e^{4\xi^2/3b^2},
\een\ees
with $g_3(x,t)$ given by \eqref{g3}. In Fig.~2(b) we plot the trapping potential \eqref{v2} in the standard form, for the resonant solution, to show how it behaves as a function of space and time. Here we see that for the darklike soliton, the trapping potential has a different behavior, if compared with the case of brightlike soliton -- see Fig.~2. The presence of a nonvanishing $\mu$ modifies both the form and deepness of the trapping potential in the darklike case.  

\begin{figure}[ht]
\includegraphics[width=4.1cm]{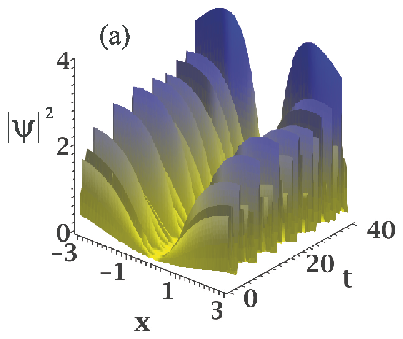}\hspace{0.1cm}
\includegraphics[width=4.1cm]{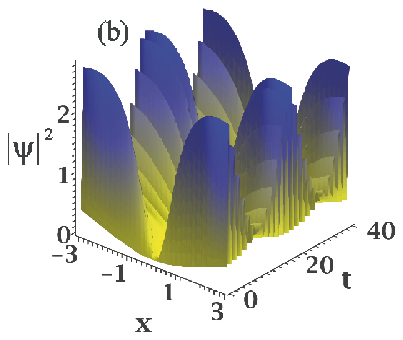}
\caption{(Color on line) Plots of $|\Psi(x,t)|^2$ for the resonant dark soliton for $\epsilon=0.5$ and $w_0=1$ (a) and of the breathing dark soliton, for $\epsilon=0$ (b). Initial data for \eqref{Ermakov} as in Fig.~3.}
\label{dsrs}
\end{figure}

In the darklike case, we also have several soliton solutions. The resonant dark soliton appears with an appropriate choice for $\varepsilon$ and $\omega_{0}$ in Eq.~\eqref{omega}, as we show in Fig.~4. We can also have the breathing soliton. It is obtained taking $\varepsilon=0$ in \eqref{omega}. In this case we have $\omega(t)=1$, and this leads us with $\gamma(t)={2}/{\sqrt{1+15\cos^{2}(2t)}}$. The $|\Psi(x,t)|^2$ for darklike soliton of the breathing type is also shown in Fig.~4. If we take $\omega _{0}=\sqrt{2}/2$ in \eqref{omega}, we can use the above $\gamma(t)$ to get to quasiperiodic solitons, as we show explicitly in Fig.~5.

Up to here we have chosen $\delta(t)\!\!=\!\!0$, to fix the center of mass of the solution; however, if we choose $\delta(t)\!\!\neq\!\!0$ we can make it to move. For simplicity, we consider the case $\varepsilon\!=\!0$ in \eqref{omega}, and take $\gamma(t)$ as above. Thus, we get to $f_1(t)\!=\!0$ in \eqref{13b} and $f_2(t)\!=\!0$ in \eqref{13c} with the introduction of $\tau(t)\!=\!\int\gamma^{2}(t)dt$, which leads to $\delta(t)\!=\!\cos(2\tau(t))$ and $a(t)\!=\!\tau(t)\!+\!(1/4)\sin(4\tau(t))$. With this, we get to the case of a moving dark soliton, as we show explicitly in Fig.~6.

\begin{figure}[ht]
\includegraphics[width=4.2cm]{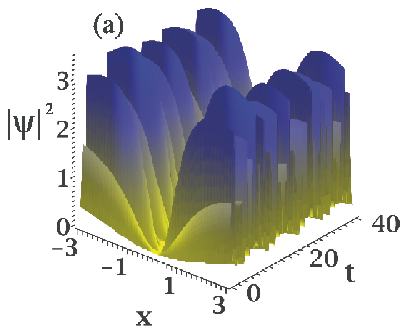}\hspace{0.2cm}
\includegraphics[width=3.4cm]{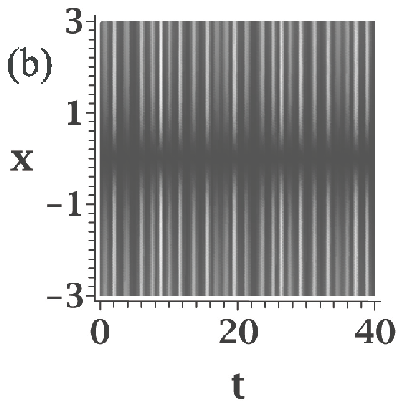}
\caption{(Color on line) Plot of $|\Psi(x,t)|^{2}$ for the quasiperiodic dark soliton with $\varepsilon=0.5$ and $\omega_{0}=\sqrt{2}/2$ in \eqref{omega} (a). The quasiperiodic behavior is better seen from the soliton profile at panel (b). Initial data for \eqref{Ermakov} as in Fig.~3.}
\end{figure}

In this work we have extended the procedure set forward in the recent Letter \cite{BB} to the case of cubic and quintic nonlinearities. As we have shown explicitly, the procedure is robust and can be generalized to higher (odd) nonlinearities very naturally, algorithmically. This result opens a new route for bright and dark solitons, with a diversity of possibilities of practical applications.

\begin{figure}[t]
\includegraphics[width=4.2cm]{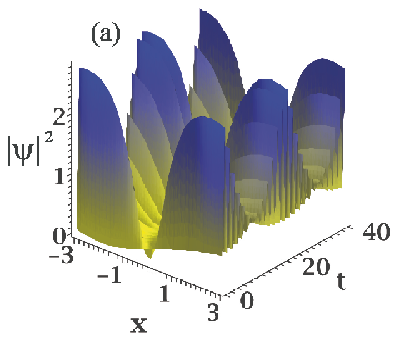}\hspace{0.2cm}
\includegraphics[width=3.4cm]{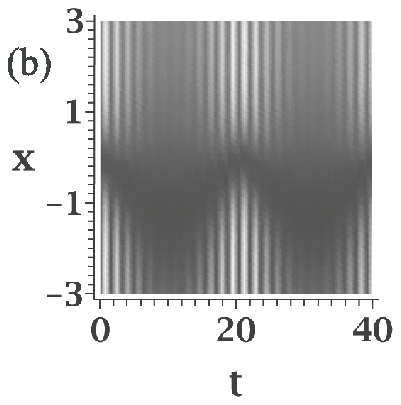}
\caption{(Color on line) Plot of $|\Psi(x,t)|^2$ for the moving breathing dark soliton (a). The center of mass motion is better seen from the profile at panel (b). Initial data for \eqref{Ermakov} as in Fig.~3.}
\end{figure}

To strengthen the result, we have shown explicitly that the system can support bright or dark solitons, depending on the way we make the eigenvalue of the stationary nonlinear Eq.~\eqref{Q} to vanish, in the case of bright soliton, or not, in the case of dark soliton. In the specific case of dark solitons, we have shown how to find the solutions with the resonant, breathing, quasiperiodic and moving behavior. A more detailed discussion is under preparation \cite{abc1}. 

The authors would like to thank CAPES, CNPQ and PRONEX-CNPq-FAPESQ for partial financial support.


\end{document}